\RequirePackage{fix-cm}
\documentclass[pdflatex,smallextended]{svjour3}       
\smartqed  
\usepackage{graphicx}
\usepackage[breaklinks=True]{hyperref}
\newcommand{\degree}{\hbox{$^\circ$$$}}
\journalname{Experimental Astronomy}
\begin{document}

\title{The Kilo-Degree Survey
}

\titlerunning{The Kilo-Degree Survey}        

\author{Jelte T. A. de Jong, Gijs A. Verdoes Kleijn, Konrad H. Kuijken, Edwin A. Valentijn,  KiDS and Astro-WISE consortiums}

\authorrunning{de Jong et al.} 

\institute{J.T.A. de Jong, K. Kuijken \at
              Leiden Observatory, Leiden University, P.O. Box 9513, 2300 RA Leiden, The Netherlands \\
              \email{jelte@strw.leidenuniv.nl}           
           \and
           G. Verdoes Kleijn, E. Valentijn \at
              OmegaCEN, Kapteyn Astronomical Institute, University of Groningen, P.O. Box 800, 9700 AV Groningen, The Netherlands
}

\date{Received: date / Accepted: date}

\maketitle

\begin{abstract}

The Kilo Degree Survey (KiDS) is a 1500 square degree optical imaging
survey with the recently commissioned OmegaCAM wide-field imager on
the VLT Survey Telescope (VST).  A suite of data products will be
delivered to the European Southern Observatory (ESO) and the community
by the KiDS survey team. Spread over Europe, the KiDS team uses
\textsf{Astro-WISE} to collaborate efficiently and pool hardware
resources. In \textsf{Astro-WISE} the team shares, calibrates and
archives all survey data.  The data-centric architectural design
realizes a dynamic 'live archive' in which new KiDS survey products of
improved quality can be shared with the team and eventually the full
astronomical community in a flexible and controllable manner.


\end{abstract}



\section{Introduction}

One of the radical advances that optical astronomy has seen in recent
years is the advent of wide-field CCD-based surveys. Key ingredients
for these surveys are the availability of instruments with
sufficiently large arrays of high quality CCD's, as well as
information systems with sufficient computing and data storage
capabilities to process the huge data flows. The scientific importance
of such surveys, particularly when freely available to researchers, is
clearly demonstrated by the impact that surveys such as 2MASS
\cite{2mass} and the Sloan Digital Sky Survey (SDSS) \cite{sdssdr8},
have had in several fields in astronomy.

Due to their telescopes being located in the Northern hemisphere, both
SDSS and the currently on-going PanSTARRS
survey \footnote{http://ps1sc.org} mainly survey the Northern sky. So
far, a similar large scale survey has not been performed from the
South and no dedicated optical survey telescope were operational until
recently. For European astronomy, however, the Southern hemisphere is
especially important, due to the presence of ESO's Very Large
Telescope (VLT) and its large array of instruments. This is now
remedied with the arrival of ESO's own two dedicated survey
telescopes: VISTA in the (near-)infrared and the VLT Survey Telescope
(VST) in the optical. Both have become operational during the past two
years. The lion's share of the observing time on both survey
telescopes will be invested in a set of `public surveys'. In terms of
observing time, the largest of the optical surveys is the Kilo-Degree
Survey (KiDS), which is imaging 1500 square degrees in four filters
($u$,$g$,$r$,$i$) over a period of 3--4 years. Combined with one of
the VISTA surveys,
VIKING\footnote{http://www.eso.org/sci/observing/policies/PublicSurveys/sciencePublicSurveys.html},
which is observing the same area in ZYJHK, this will provide a
sensitive, 9-band multi-colour survey.

Specifically for the handling of surveys from the VST the Astro-WISE
system \cite{astrowise} has been designed. It allows processing,
quality control and public archiving of surveys using a distributed
architecture. The KiDS survey team, that is spread over different
countries, performs these survey operations in Astro-WISE as a single
virtual team making intensive use of web-based collaborative
interfaces. This paper will discuss both the observational set-up of
the KiDS survey and its primary scientific goals, as well as how the
Astro-WISE system will be used to achieve these goals.

\section{The Kilo-Degree Survey}

\subsection{Instrumental set-up}

The VST is located at Paranal Observatory in Chile and operated by
ESO. Regular observations with the system commenced on October 15th
2011. With a primary mirror of 2.6-m diameter it is currently the
largest telescope in the world specifically designed for optical
wide-field surveys. The sole instrument of the VST is OmegaCAM
\cite{omegacam1}, a 268 Megapixel wide-field camera that provides a
1\degree$\times$1\degree~field-of-view. The focal plane array is built
up from 32 2048$\times$4096 pixel CCDs, resulting in 16k$\times$16k
pixels with a pixel scale of 0.214 arcseconds/pixel. The optics of the
telescope and camera were designed to produce a very uniform
point-spread-function over the full field-of-view, both in terms of
shape and size. For a more detailed description we refer the reader to
the OmegaCAM paper in this issue \cite{omegacam2}.

\begin{figure}[ht]
\includegraphics[width=\textwidth]{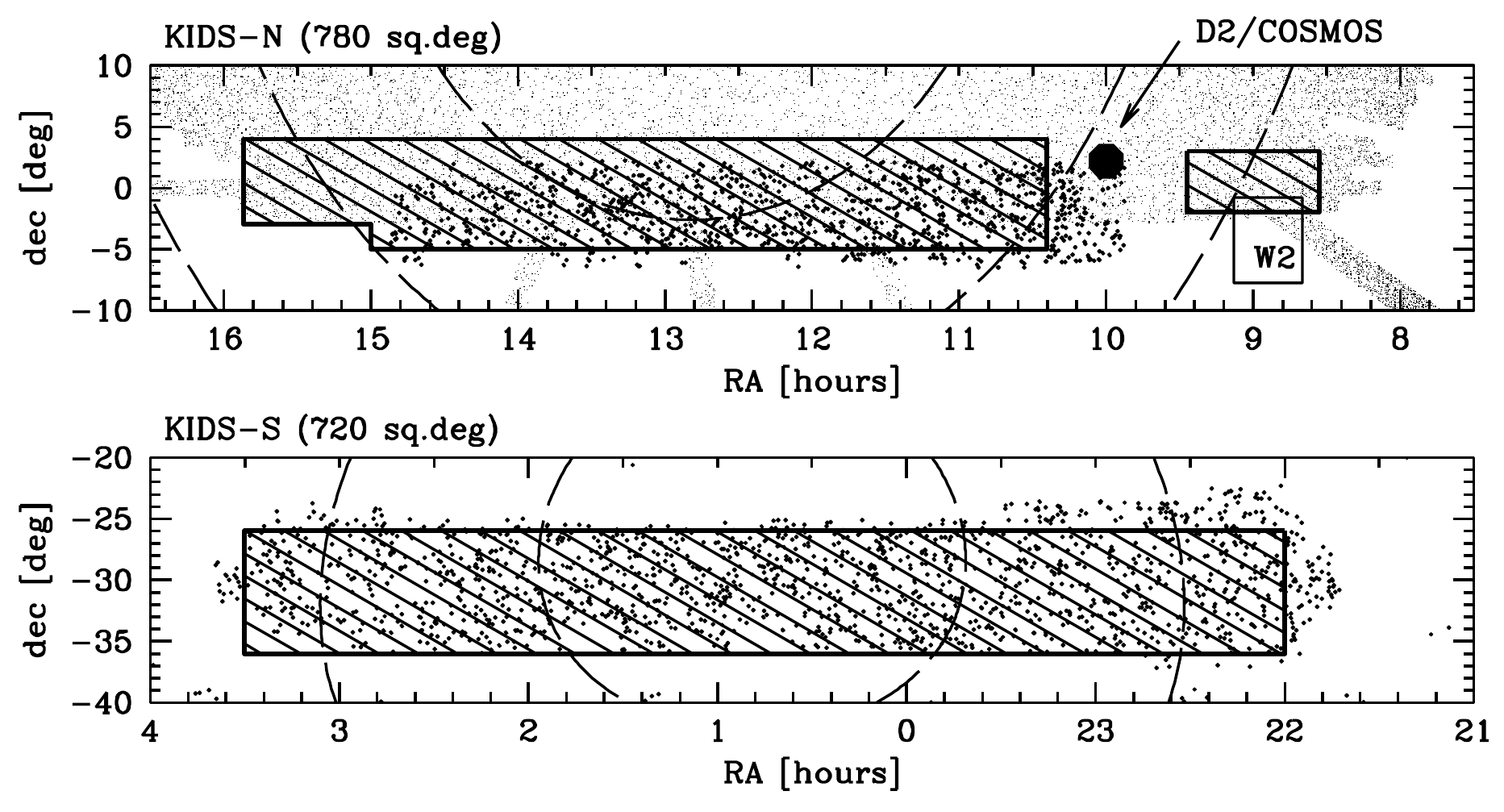}
\caption{Lay-out of the KIDS-North (top) and KIDS-South (bottom)
  fields, shown by the hatched areas. Also shown are the areas where
  2dF spectra are available, indicated by the large dots, and the area
  covered by DR7 of the SDSS survey, indicated by the small dots. The
  CFHTLS-W2 field and the DS/COSMOS deep field are overplotted on the
  top panel.}
\label{fig:areas}
\end{figure}

\begin{table}[ht]
\caption{KiDS Fields}
\label{tab:fields}
\begin{tabular}{l|lll}
\hline\noalign{\smallskip}
Field & RA range & Dec range & Area \\
\noalign{\smallskip}\hline\noalign{\smallskip}
KiDS-S    & 22$^h$00$^m$ -- 3$^h$30$^m$ & $-$36\degree -- $-$26\degree & 720 sq.deg \\
\noalign{\smallskip}\hline\noalign{\smallskip}
KiDS-N    & 10$^h$24$^m$ -- 15$^h$00$^m$ & $-$5\degree -- $+$4\degree & 712 sq.deg \\
~         & 15$^h$00$^m$ -- 15$^h$52$^m$ & $-$3\degree -- $+$4\degree & ~ \\
\noalign{\smallskip}\hline\noalign{\smallskip}
KiDS-N-W2 & 8$^h$30$^m$ -- 9$^h$30$^m$ & $-$2\degree -- $-$3\degree & 68 sq.deg \\
\noalign{\smallskip}\hline\noalign{\smallskip}
KiDS-N-D2 & 9$^h$58$^m$ -- 10$^h$02$^m$ & $+$1\degree -- $+$3\degree & 2 sq.deg \\
\noalign{\smallskip}\hline
\end{tabular}
\end{table}

\subsection{Observational survey set-up}

KiDS will cover 1500 square degrees, some 7\% of the extragalactic
sky. It consists of two patches, ensuring that observations can take
place year-round. The Northern patch lies on the celestial equator,
while the Southern patch straddles the South Galactic Pole; see
Fig. \ref{fig:areas} and Table \ref{tab:fields} for the detailed
lay-out. Together the two patches cover a range of galactic latitudes
from 40 to 90 degrees, and the 10 degree width of the strips ensures
that the full 3D structure of the universe is sampled well. These
specific areas were chosen because they have been the target of
massive spectroscopic galaxy surveys already: the 2dF redshift survey
\cite{2dfgrs} covers almost the same area, and KiDS-N overlaps with
the SDSS spectroscopic and imaging survey as well. This means that
several 100,000 galaxy spectra and redshifts are already known in
these fields, and hence that the cosmological foreground mass
distribution in these fields is well mapped out. Extinction in the
fields is low.  The exposure times for KiDS and VIKING have been
chosen to yield a median galaxy redshift of 0.8, so that the evolution
of the galaxy population and matter distribution over the last $\sim$
half of the age of the universe can be studied. They are also
well-matched to the natural exposure times for efficient VST and VISTA
operations, and balanced over the astro-climate conditions on Paranal
(seeing and moon phase) so that all bands can be observed at the same
average rate. This strategy makes optimal use of the fact that all
observations are queue-scheduled, making it possible to use the best
seeing time for deep $r$-band exposures, for example, and the worst
seeing for $u$. All exposure times and observing constraints are
listed in Table \ref{tab:exptimes}, where seeing refers to the
full-width-half-maximum (FWHM) of the point-spread-function (PSF)
measured on the images.

\begin{table}
\caption{KiDS Exposure Times and Observational Constraints}
\label{tab:exptimes}
\begin{tabular}{lcccccc}
\hline\noalign{\smallskip}
Filter & Exposure time & Mag limit & PSF FWHM & Moon & ADC & Airmass \\
  & (seconds) & (AB 5$\sigma$ 2'') & (arcsec) & phase & used & \\
\noalign{\smallskip}\hline\noalign{\smallskip}
$u$ & 900 & 24.8 & 0.9--1.1 & Dark & no & 1.2 \\
$g$ & 900 & 25.4 & 0.7--0.9 & Dark & no & 1.6 \\
$r$ & 1800 & 25.2 & $<$0.7 & Dark & no & 1.3 \\
$i$ & 1080 & 24.2 & $<$1.1 & Any & no & 2.0 \\
\noalign{\smallskip}\hline
\end{tabular}
\end{table}

Since the OmegaCAM CCD mosaic consists of 32 individual CCDs, it is
not contiguous but contains gaps. To avoid holes in the KiDS images,
observations will use 5 dithered observations per field in $g$, $r$
and $i$ and 4 in $u$.  The dithers form a staircase pattern with
dither steps of 25'' in X and 85'' in Y. These offsets bridge the
inter-CCD gaps of OmegaCAM.  The survey tiling is derived using a
tiling strategy that can tile the full sky efficiently for the
OmegaCAM instrument. Neighboring tiles have an overlap in RA of 5\%
and in DEC of 10\%. This will allow us to derive the photometric and
astrometric accuracies required for the most stringent science cases:
internal astrometric error $<0.03''$ rms and 1\% photometric
errors. The Atmospheric Dispersion Corrector (ADC) of OmegaCAM could
be used for all KiDS bands except $u$. However, KiDS does not make use
of the ADC to avoid the small losses in sensitivity. Instead, the
dispersion is limited by constraining the maximum airmass (see Table
\ref{tab:exptimes}). Particularly for $r$ this is important since this
band will be used for weak lensing analyses and therefore requires a
well-behaved PSF. By constraining the maximum airmass in $r$ to 1.3,
the spectral dispersion will be $<$0.2''.

After completion of the main survey of 1500 square degrees, the whole
survey area will be imaged once more in $g$-band. The observational
set-up of this repeat pass is the same as for the main survey
observations in $g$, with the additional requirement that it should
provide at least a 2 year baseline over the whole survey area. With
the average $g$-band seeing of 0.8'' this will allow for proper motion
measurements with accuracies of 40 mas yr$^{-1}$ and better for
sources detected at signal-to-noise of 10 \cite{kuijkenrich02}.

\subsection{Scientific goals}

The central science case for KiDS and VIKING is mapping the matter
distribution in the universe through weak gravitational lensing and
photometric redshift measurements. However, the enormous data set that
KiDS will deliver, will have many more possible applications. The main
research topics that the KiDS team members will explore are outlined
below.

\subsubsection{Dark energy}

Dark energy manifests itself in the expansion history of the universe,
as a repulsive term that appears to behave like Einstein's
cosmological constant. Understanding its properties more accurately is
one of the central quests of cosmology of recent years
(e.g. \cite{zhao10}). With KiDS we intend to push this question as far
as possible, while recognising that the limiting factor may well be
systematic effects rather than raw statistical power. Measurements
done with KiDS will therefore also serve as a learning curve for
future (space-based) experiments.

\begin{figure}[t]
\includegraphics[width=\textwidth]{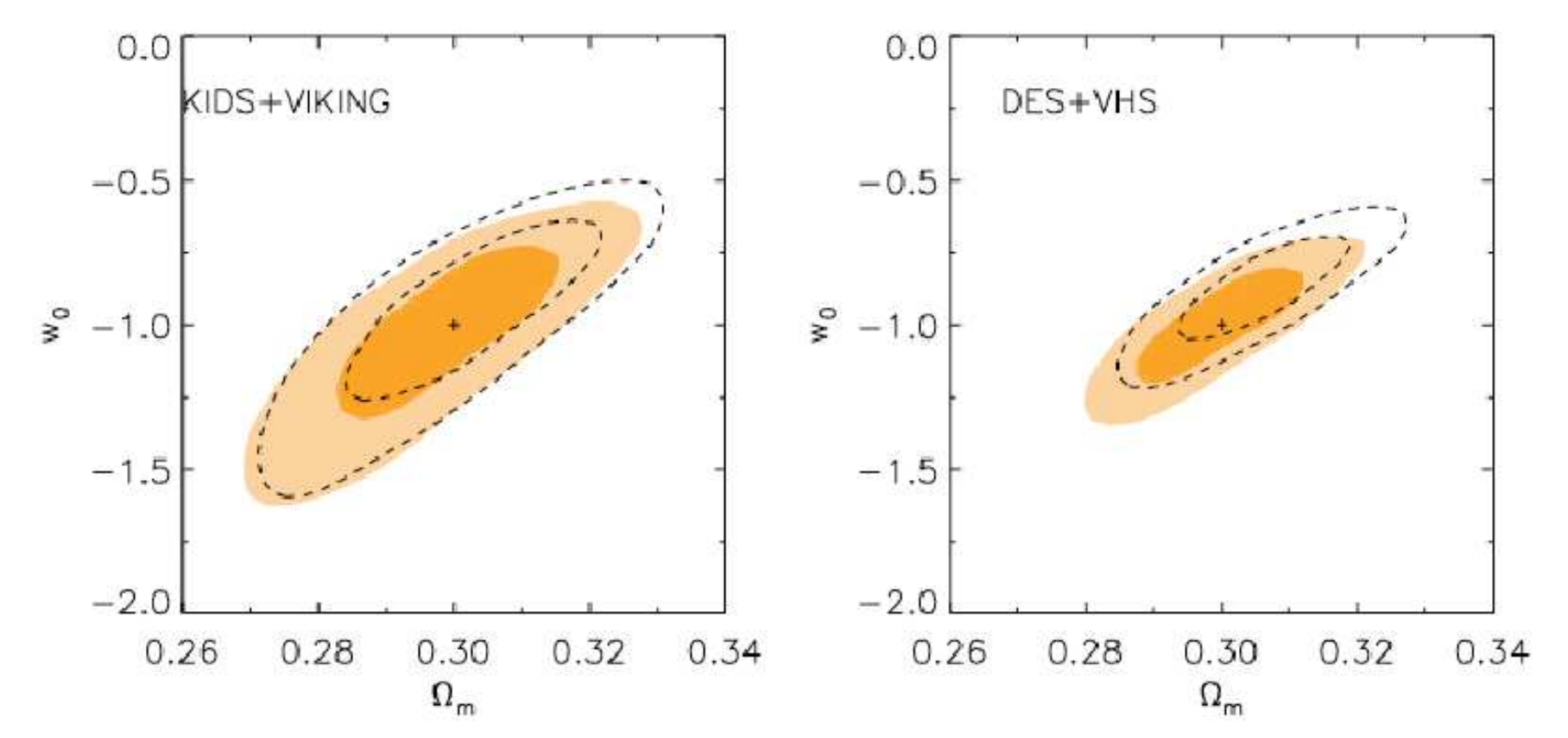}
\caption{Comparison of the formal statistical power and sensitivity to systematic errors in the photometric redshifts for 
the KiDS/VIKING and DES/VHS surveys in cosmological parameter estimation (here the matter density $\Omega_m$ and the dark 
energy equation of state $w_0$), based on simulated photometry for each of the surveys \cite{szomoru12}. 
The '+' represents the input truth. The coloured contours assume perfect redshift information, while the dashed contours show the 
effect of redshift errors. Flat geometry was assumed here, but otherwise no external information were included. Once external 
information is folded in, the constraints tighten and systematic effects become even more significant, demonstrating 
the greater robustness of the KiDS survey to this type of systematic error.}
\label{fig:cosmo_parameters}
\end{figure}

The use of weak gravitational lensing as a cosmological probe is
nicely summarized in \cite{mellier99} and
\cite{peacock06}. Essentially, its power relies on two facts:
gravitational lensing is a very geometric phenomenon, and it is
sensitive to mass inhomogeneities along the lines of sight.This makes
it a good probe of the growth of structure with time (redshift), as
well as being a purely geometric distance measure. As it happens, the
distance-redshift relation and the speed with which overdensities grow
with cosmic time are the two most fundamental measures of the energy
content of the universe: both depend directly on the rate at which the
universe expands. Making such a measurement, for which weak lensing is
an excellent method, is therefore of great interest.  These lensing
measurements are not easy, as they require systematics better than 1\%
accuracy, and photometric redshifts unbiased at a similar
level. However, with KiDS we have put ourselves in the optimal
position to attempt this, by ensuring the best image quality in our
instrument, by choosing a survey depth and area appropriately, and
having a wide wavelength coverage that will make the photometric
redshifts as free of error as is possible with wide-band photometry
(see Fig. \ref{fig:cosmo_parameters}).  As noted above, the expansion
history can be deduced from lensing tomography in several ways, and
requiring consistency is a powerful check -- as well as, further down
the line, an interesting test of Einstein gravity theory
\cite{jain08}.

An independent way to study the expansion history of the universe is
by measuring the baryon acoustic oscillations (BAO). BAO is the
clustering of baryons at a fixed co-moving length scale, set by the
sound horizon at the time that the universe recombined and photons
decoupled from baryonic matter. This scale length, which has been
measured accurately in the cosmic microwave background
\cite{spergel07}, is therefore a standard ruler, whose angular size on
the sky provides a direct measurement of the angular diameter-redshift
relation and hence of the expansion history. Using photometric
redshifts from KiDS we can make an independent measurement of the BAO
scale. Comparison of the results with recent and ongoing spectroscopic
BAO surveys such as WiggleZ \cite{wigglez} and BOSS \cite{boss}
provides a potent test of systematics. Simulations of photometric
redshift measurements using full 9-band coverage, ugrizYJHK as will be
provided by KiDS and VIKING, have shown that the accuracy needed for
detection of the BAO signal ($0.03\times(1+z)$ rms photometric
redshift error) can be reached.  Tests of the detectability of the BAO
with particle and Monte-Carlo simulations, provided by Peter
Schuecker, have shown that imaging surveys of the size and sensitivity
of KiDS can yield values of $w$ with $\sim$5\% accuracy.

\subsubsection{Structure of galaxy halos}

Simulations of structure formation provide detailed information about
the shape of dark matter halos on large scales. However, at small
scales such as the inner parts of galaxy halos, complex physics that
these simulations can not represent realistically (e.g. star
formation, cooling, feedback etc.) starts to play an important role
(e.g. \cite{vandaalen11}). The relation between light (baryons) and
mass (dark matter) is crucial for our understanding of the influence
of the dark matter on galaxy formation, and vice-versa. Galaxy-galaxy
lensing (GGL) provides a unique way to study this relation between
galaxies and their dark halos.

The gravitational lensing effect of foreground galaxies on the images
of background galaxies is very weak, and can only be measured
statistically. This is done by stacking large numbers of foreground
galaxies and measuring the net image distortion of the background
galaxies. On small scales (3--30 arcsec) the signal is dominated by
the profile of the foreground galaxies' inner dark matter halos, at
radii of 10 to 100s of kpc. At scales of several arcminutes GGL probes
the galaxy--mass correlation and the bias parameter, while at even
larger scales the distribution of the foreground galaxies in their
parent group halos dominates. GGL can therefore be used to probe halos
over a large range of scales and help to test the universality of the
dark halo profile.

The strength of KiDS for GGL is again twofold. The shear size and thus
enormous numbers of available galaxies, makes it possible to split the
foreground galaxies in bins and study different galaxy types
separately. The accurate photometric redshifts also allow splitting up
the samples in redshift bins, thus enabling the redshift dependence to
be analyzed. Furthermore, the fact that KiDS targets areas where
wide-field redshift surveys have already been carried out, means that
the foreground large scale structure is known, enabling the
measurement of the galaxy--mass correlation for galaxy groups,
clusters, and even filaments.  Compared to earlier GGL studies with,
for example, SDSS \cite{mandelbaum06} or CFHTLS
\cite{parker07}, the image quality and sensitivity of KiDS
will provide many more foreground--background pairs, more accurate
shape measurements, and the ability to probe the galaxy population at
higher redshifts.

\subsubsection{Evolution of galaxies and clusters}

Within the current cosmological paradigm of Cold Dark Matter (CDM),
structure formation is hierarchical and the profiles of CDM halos are
universal, i.e. the same at all scales. Several of the ramifications
of this picture have so far eluded rigorous observational testing. For
example, various observational constraints on the influence of galaxy
mergers on the evolution of the galaxy population at redshifts higher
than $\sim$0.5 differ up to an order of magnitude (see
e.g.~\cite{mergersLotz11}). The observational studies targeted small
numbers of galaxies ($<1000$) at high spatial resolution (e.g.,
\cite{mergersMan12}) or small areas ($<1$ square degree,
e.g. \cite{mergersCooper12}). Also, galaxy clusters are probes of the
highest mass peaks in the universe, but at redshifts of $z>$1 the
number of known galaxy clusters is yet too small to constrain
cosmological models.

KiDS can play a major role in this field. The sensitivity of the KiDS
photometry will result in the detection of an estimated $10^8$
galaxies. This galaxy sample will have a median redshift of $z=0.8$,
with $\sim$20\% having $1<z<1.5$. Based on this sample the evolution
of the galaxy luminosity function, the build-up of stellar mass and
the assembly of early-type stellar systems can be traced back to
unprecedented look-back times. 

Cluster finding will be possible directly from the multi-colour KiDS
catalogues. In total we expect KiDS to provide $1-2\times10^4$
clusters, and with the red sequence detectable out to $z\sim1.4$
approximately 5\% of these will be located at redshifts beyond 1. This
will be a very important sample to further constrain cosmological
parameters, provided that the relation between cluster richness and
cluster mass can be calibrated. This calibration is possible since the
weak lensing measurements that will be done as part of KiDS will probe
the cluster mass distribution, demonstrating the pivotal advantage of
combining high image quality with uniform multi-band photometry.

A different perspective of galaxy evolution will be provided by virtue
of the fact that that KiDS-S overlaps two nearby superclusters
(Pisces-Cetus and Fornax-Eridanus). Thus, the relation between galaxy
properties (e.g. star formation rate) and environment, can be
studied all the way from cluster cores to the infall regions, and to
the filaments that connect clusters in the cosmic web.

\subsubsection{Stellar streams and the Galactic halo}

Detailed studies of the stellar halo of the Milky Way require
photometry of faint stars over large areas of sky. The SDSS, although
primarily aimed at cosmology and high-redshift science, has proved a
milestone in Milky Way science as well, unveiling many stellar streams
and unknown faint dwarf Spheroidal galaxies
\cite{fieldofstreams,catsanddogs}. While KiDS will image a
smaller area than SDSS, it is deeper and thus will provide a view on
more distant parts of the halo. But more importantly, SDSS only
covered the Northern sky, leaving the Southern hemisphere as uncharted
territory. Particularly in the KiDS-S area, new discoveries are bound
to be made in the direct vicinity of our own Galaxy.

\subsubsection{Proper motions}

Proper motions with accuracies of $\sim0.04 ''yr^{-1}$ will be
available in the KiDS area, owing to the planned g-band repeat pass
that will provide a 2-year baseline. Several applications are
possible, among others the detection and study of high proper-motion
white dwarfs. ``Ultracool'' white dwarfs ($<$4000 K), relics from the
earliest epochs of star formation, are among the oldest objects in the
Galaxy and can be used to trace the very early star formation history
of our Galaxy. Due to its multicolour photometry combined with proper
motion information, KiDS will be able to increase the sample of known,
ultracool white dwarfs \cite{harris08,kilic10} significantly.

\subsection{Survey calibration and data products}

Being a Public Survey, all KiDS data will be made publicly
available. The KiDS catalogue will contain some 100,000 sources per
square degree (150 million sources over the full survey area), and for
each square degree there will be 10GB of final image data, 15TB for
the whole survey. These data will be of no use if they are not
uniformly and carefully calibrated and made available in an easily
accessible archive. 

The astrometric calibration is done per tile (stack of dithers) using the gnomonic tangential projection and using the 2MASS Point Source Catalog (2MASS PSC, \cite{2mass}) as survey astrometric reference. The first step is a local astrometric solution. An astrometric model for a single chip is constrained by the star positions from a single exposure. The local solution is the input for global astrometry. The global solution uses a global model of the focal plane that allows for variations over the dither positions that make up the tile (e.g., due to telescope flexure and field rotation tracking residuals). The observational constraint comes from the internal positional residuals from dither overlaps and external residuals with the 2MASS PSC. First results for local astrometry indicate $0.1"-0.15"$ rms in relative astrometry for KiDS. Global astrometry is expected to yield $\sim 0.05"$ rms.

The photomeric calibration will be done over the survey as a whole in Sloan photometric system using AB magnitude system. The Calibration Plan of the survey telescope includes nightly zeropoints on SA standard fields in u, g, r and i. A dedicated OmegaCAM calibrational observing program is obtaining Secondary Standards in the SA fields covering the full OmegaCAM FoV. The program is expected to run until the last quarter of 2012. Each night $\sim$three observations are taken of a fixed polar standard field near the southern equatorial pole for atmospheric monitoring using a composite filter with u, g, r and i quadrants. Daily domeflats are taken in all bands to measure the system throughput (telescope plus camera). Commissioning results indicate that the on-sky observations yield photometric scales to $< 1\%$ accuracy and the in dome observations an $<0.5\%$ accuracy. The overlaps between KiDS science observations constrain the relative photometric calibration over the full survey. For the absolute photometric calibration the standard field observations are used. The fellow ATLAS survey\footnote{http://astro.dur.ac.uk/Cosmology/vstatlas/} using OmegaCAM on the VST has full overlap with the KiDS-S area and $\sim 25\%$ overlap with the KiDS-N area in u, g, r and i. Tying the KiDS survey to these ATLAS areas plus the standard fields shall prevent calibration "creep" from the tile-to-tile photometric calibration. \cite{opticalPipeline} give a detailed description of the Astro-WISE calibration pipeline. The OmegaCAM and VST calibration is discussed in detail in \cite{omegacam2}. They also discuss the instrument photometric characterization, including illumination variations.  

Basic data products will be made public, both
through ESO and through the Astro-WISE database, within a year after
any part of the survey area has been observed in all filters. This set
of basic data products includes the following:
\begin{itemize}
\item astrometrically and photometrically calibrated coadded and regridded images with weight maps;
\item calibration images: twilight flats, dome flats, biases, fringe maps, etc.;
\item single-band source catalogues;
\item multi-color (i.e. combined single-band) catalogues.
\end{itemize}

In addition, and on a longer time-scale, we intend to provide more
refined and advanced data products. In the context of the lensing
project for KiDS several innovative image processing techniques have
been developed, and to the extent possible these will be used to
generate high-level data products in the KiDS database. Many of the
parameters developed for the SDSS survey will be provided.
Furthermore we will look at including:

\begin{itemize}
\item Images with ''Gaussianized'' PSF. Versions of all images convolved
  with kernels chosen to result in a homogenized, round and
    Gaussian PSF, to ease comparison with images taken at different
  times or with different filters or instruments.
\item Aperture-matched colour catalogues. Catalogues with colours
  measured only from the high S/N inner regions of sources, for
  applications that only require flux ratios, rather than total
  fluxes.
\item Unsharp masked images. A wealth of underlying galaxy structure
  can be obtained from images in which the low frequencies have been
  removed (dust structures, disks, etc.). We plan to provide images
  filtered in various ways.
\item Morphological parameters. The popular galaxy profile fitting
  programmes GALFIT \cite{peng02} and GALPHOT \cite{franx89} have been
  implemented in AstroWISE, and will be run on the sources and
  published in the KiDS database.
\end{itemize}

Although the large-scale data processing for the VIKING survey will
not be done by the KiDS team, certain VIKING data products, and
combined KiDS-VIKING data products will be made available through
Astro-WISE. For a more extensive discussion on how VIKING data
products will be ingested and processed (where needed) within
Astro-WISE, see the Data Zoo paper in this issue \cite{datazoo}.

\section{Internet-based survey collaboration for a geographically distributed team}

The KiDS survey team is an international collaboration with $\sim 35$
team members at institutes spread around Europe and beyond. The team
uses Astro-WISE for KiDS survey handling: data processing (image
calibration, stacking, cataloging), data quality control and data
management (on-line archiving and publishing in the Virtual
Observatory). By logging on to a single system, Astro-WISE, team
members can make use of a distributed pool of storage, compute and
database resources spread over Europe, and do their survey work,
irrespective of where this person is. Day-to-day survey handling is
done via webservices. Thus, a web browser with internet connection is
all that is required to start doing KiDS survey work.

\subsection{Astro-WISE: a data-centric approach}

This 'whoever, wherever' approach is possible because all aspects of
survey handling in Astro-WISE are implemented from a data-centric
viewpoint. For example, calibration scientists add information on the
time-validity of a calibration item (e.g. a master flatfield image) to
this item: the information {\it becomes part of} the data item
itself. Quality control is handled similarly, since the verdicts of
both automatic and human quality assessors become part of the data
items themselves. The same is true for data management, because rather
than users knowing which data they can reach, in Astro-WISE each data
item 'knows' (contains information on) which users it can
reach. Processing is implemented as data reaching compute clusters,
not as users reaching compute clusters, and all survey data can access
all hardware that is pooled by the KiDS team.

\begin{figure}[ht]
\includegraphics[width=11.9cm]{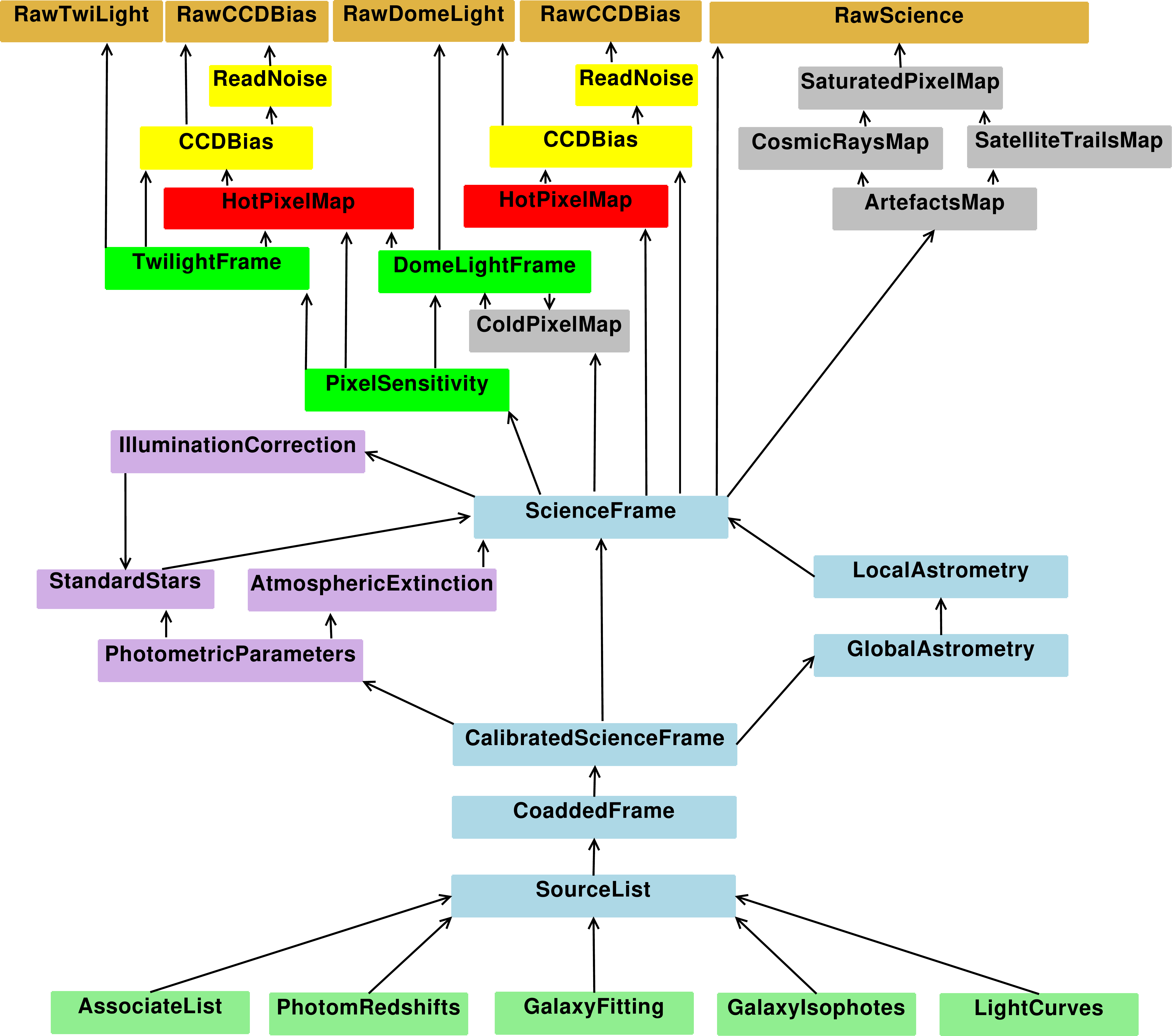}
\caption{Overview of classes of data objects. Data objects not only
  contain the survey products denoted by familiar names in wide-field
  imaging. They also carry the information how they, as requested
  target, can be created out of other survey objects, illustrated by
  the arrows. Underlying is an object model that captures the
  relationship between requested information and the physics of the
  atmosphere-to-detector observational system.}
\label{f:targetDiagram}
\end{figure}

Moreover, a survey data item also 'knows' how it was made, and whether
it could be made better using new, improved calibrations. This is
possible because all processing operations are implemented as actions
by data objects acting upon themselves and/or other data objects.
Each type of survey product, from raw to final, is represented by a
class of data objects, and each survey product is a data object: an
informational entity consisting of pixel and/or metadata, where
metadata is defined as {\it all} non pixel data.  Final survey
products also carry the information on how they can be created out of
intermediate survey objects. This backward chaining procedure is
recursively implemented up to the raw data (see
Figure~\ref{f:targetDiagram} and \cite{datalineage}). Thus, a request
by a KiDS team member for a survey product, a target, triggers a
backward information flow, in the direction of the raw data. The net
effect is a forward work flow description, towards the target, that is
then executed. The backward information flow is implemented as queries
to database initiated by the requested target itself. The database is
queried for survey objects on which the target depends with the right
characteristics including validity and quality. Either they exist and
are returned or the query is 'backwarded' to the next level of survey
objects closer to the raw data.

In conclusion, Astro-WISE takes a data-centric approach to survey
handling and control. Attributes of data objects solely determine
which calibration data are applied to which science data, which survey
products have been qualified and which products should be considered
experimental or baseline. Data processing is realized by backward
information flows that results in a forward processing flow. See the
WISE paper in this special issue \cite{wise} for more information on
Astro-WISE itself.

\subsection{Managing the KiDS survey data} 

Handling of KiDS survey data (and any dataset in Astro-WISE) is based
on a number of parameters and data attributes that regulate which
users can interact with which data objects, and contain information on
the quality and validity of data objects. Now follows a detailed
description of these paramaters, summarized in
Table~\ref{t:attributesForSurveyControl}.

\begin{table}
\caption{Set of data-item attributes used for data management, calibration control, quality assessment and control in Astro-WISE.}
\label{t:attributesForSurveyControl}
\begin{tabular}{lp{4.5cm}p{4cm}}
\hline\noalign{\smallskip}
attribute & content description & possible values \\
\noalign{\smallskip}\hline\noalign{\smallskip}
\_creator & Astro-WISE user that created the data object & any Astro-WISE user name \\
\_project & project data object belongs to & any Astro-WISE project name \\
\_privileges & operational level at which data object resides & 1,2,3,4,5 \\
is\_valid & validity indicator set by user & 0(bad),1(no verdict),2(good) \\
quality\_flags & quality flag set by system & any integer (bitwise), with 0$==$good \\
timestamp\_start &  start and end of validiy range in time for a calibration object \\
timestamp\_end &  \\
creation\_date & time of creation of data object \\
\noalign{\smallskip}\hline
\end{tabular}
\end{table}

\begin{enumerate}
\item {\bf Creator.} Each data object is associated with a single
  Astro-WISE user, its Creator, the user that created/ingested the
  data object. An Astro-WISE user is a person with an Astro-WISE
  database account consisting of an id number and name. Each person
  has only one account and therefore a single identity within the
  Astro-WISE system. Once created, the creator of a data object
  cannot be changed.

\item {\bf Project.} A project in Astro-WISE is a group of Astro-WISE
  users that share a set of data objects\footnote{for an overview of
    all projects see http://process.astro-wise.org/Projects}. A
  project has a project id, a name, a description, project members and
  optionally an instrument. One or more users can be member of a
  project, and a user can be member of more than one project. Each
  data object belongs to one and only one project, which is chosen
  upon the creation/ingestion of the data entity and can not be
  changed after that. Some projects have all Astro-WISE users as
  members (public projects), while other projects have a subset of
  Astro-WISE users as members (private projects). The 'KIDS' project
  is a private project and contains all data objects resulting
  from processing of KiDS survey data.

\item {\bf Privileges.}  Survey data management is facilitated by
  having pools of data at five different levels named Privileges
  levels.  The term privileges stems from the data-centric viewpoint
  of Astro-WISE. Each data object has a \_privileges attribute that
  defines its Privileges level. An object has increasing privileges to
  access users with numerical increase of its Privileges
  level. Table~\ref{tab:privileges} lists the five levels and which
  users a data object can reach at each level. The initial privileges
  of a data object are set by the creator upon the creation/ingestion
  of the data entity. This can be changed later by creator and
  project managers, a process called publishing.

\item {\bf is\_valid.}  This data attribute is the validity indicator
  as set by users. It stores the quality assessment performed by a
  survey team member.  Its default value upon creation of a data item
  is is\_valid$=1$, meaning no user assessment has taken place. The
  team member can change this to is\_valid$=0$, meaning bad quality,
  or to is\_valid$=2$, meaning data is qualified as good.

\item {\bf quality\_flags.}  This data attribute collects the quality
  flags as set by the system. Automatically the quality of objects
  (i.e., survey products) is verified upon creation. If the quality is
  compromised the quality\_flags are set to a value $\neq 0$. It is a
  bitwise flag. Each type of object (raw science frames, calibrated
  science frames, astrometric solutions) has its own definition of
  what each bit means. 

\item {\bf timestamp\_start, timestamp\_end.} These attributes of a
  calibration data item define the range in time for which it is
  applicable.  Calibration data get default timestamp ranges upon
  creation. These can be modified by survey team members. For example,
  it might be decided that a zeropoint might apply to one or more
  nights, or just a few hours instead. 

\item {\bf creation\_date.} Upon creation every object that results
  from processing has an attribute that stores the moment of its
  creation. This information is relevant as ``Newer is better'' is a
  general rule for objects to determine which calibration data item
  should be applied to them from the pool of applicable calibration
  data items.
\end{enumerate}

Handling of KiDS survey data starts by filtering the pool of available
data on which the survey handling should act. This is called setting a
'Context' in Astro-WISE, and is done using the parameters 'Project'
and 'Privileges' described above. After logging into Astro-WISE, the
KiDS team member selects the Project 'KIDS' and a minimum Privileges
level of 1 or 2. All results the member produces will be part of the 'KIDS'
Project, and the member is able to see all data available within 'KIDS'.

The baseline KiDS survey products are all part of the Project 'KIDS'
and reside at privileges level 2 (named PROJECT). These data can be
accessed only by KiDS survey team members. Each KiDS team member can
experiment in her/his own privileges level 1 (MYDB) to create improved
versions of these baseline products. Only the single team member can
access survey data at MYDB level and promote it to the PROJECT
level. A KiDS project manager can promote baseline survey data from
privileges level 2 to all higher levels. Survey data at privileges
level 3 (ASTRO-WISE) can be accessed by all Astro-WISE users. At
privileges level 4 (WORLD) the data become publicly accessible, that
is to users without an Astro-WISE account (anonymous users).  Finally
at priveleges level 5 (VO), the data are accessible also from the
Virtual Observatory. 

Thus, it is the combination of Context parameters user, Project and
Privileges level that determine what data is filtered to be accessible
for survey handling. Data handling itself is done either by using a
command-line interface (CLI) or via webservices. How-to's and
tutorials for using the CLI can be found
online\footnote{http://www.astro-wise.org/portal/aw\_howtos.shtml};
and for a description of available webservices we refer to the paper
on Astro-WISE interfaces in this issue \cite{userinterfaces}.

\begin{table}
\caption{Sharing KiDS data in Astro-WISE: privilege levels}
\label{tab:privileges}
\begin{tabular}{ll}
\hline\noalign{\smallskip}
privileges level & data is shared with \\
\noalign{\smallskip}\hline\noalign{\smallskip}
1: MYDB & only the creator \\
2: PROJECT & every member of the project to which the data object belongs \\
3: ASTRO-WISE & all Astro-WISE users \\
4: WORLD & the world: Astro-WISE users and persons \\
 & without an Astro-WISE account (latter via webservice DBViewer) \\
5: VO & the whole world and data also available through \\
 & the Virtual Observatory webservices \\
\noalign{\smallskip}\hline
\end{tabular}
\end{table}

\subsection{Quality assessment and control}

Three types of quality assessment are facilitated in Astro-WISE: 
\begin{enumerate}
\item{{\bf ``verify''}: automatic verification by the system upon creation of a data object }
\item{{\bf ``compare''}: comparison of a processing result to an earlier similar result}
\item{{\bf ``inspect''}: manual inspection by a human user of a data object}
\end{enumerate}

Although this approach is generic, and all classes of data objects
resulting from processing (i.e., ``processing targets'') have the
three types of methods implemented, the actual content of the method
is specific for the type of data. The result of the application of a
method is stored as a data attribute, namely either the quality\_flags
or the is\_valid attribute discussed in the previous
section. Additional quality control information can be stored in a
Comment object as a free string that links to the process target. The
quality of an end product depends on the pipeline configuration and
the quality of (calibration) data at all intermediate processing
stages. This generic approach to quality assessment together with the
data lineage in Astro-WISE (for details see the paper on data lineage
in this issue \cite{datalineage}) facilitates tracing back these
dependencies. 

\begin{figure}[ht]
\includegraphics[width=11.9cm]{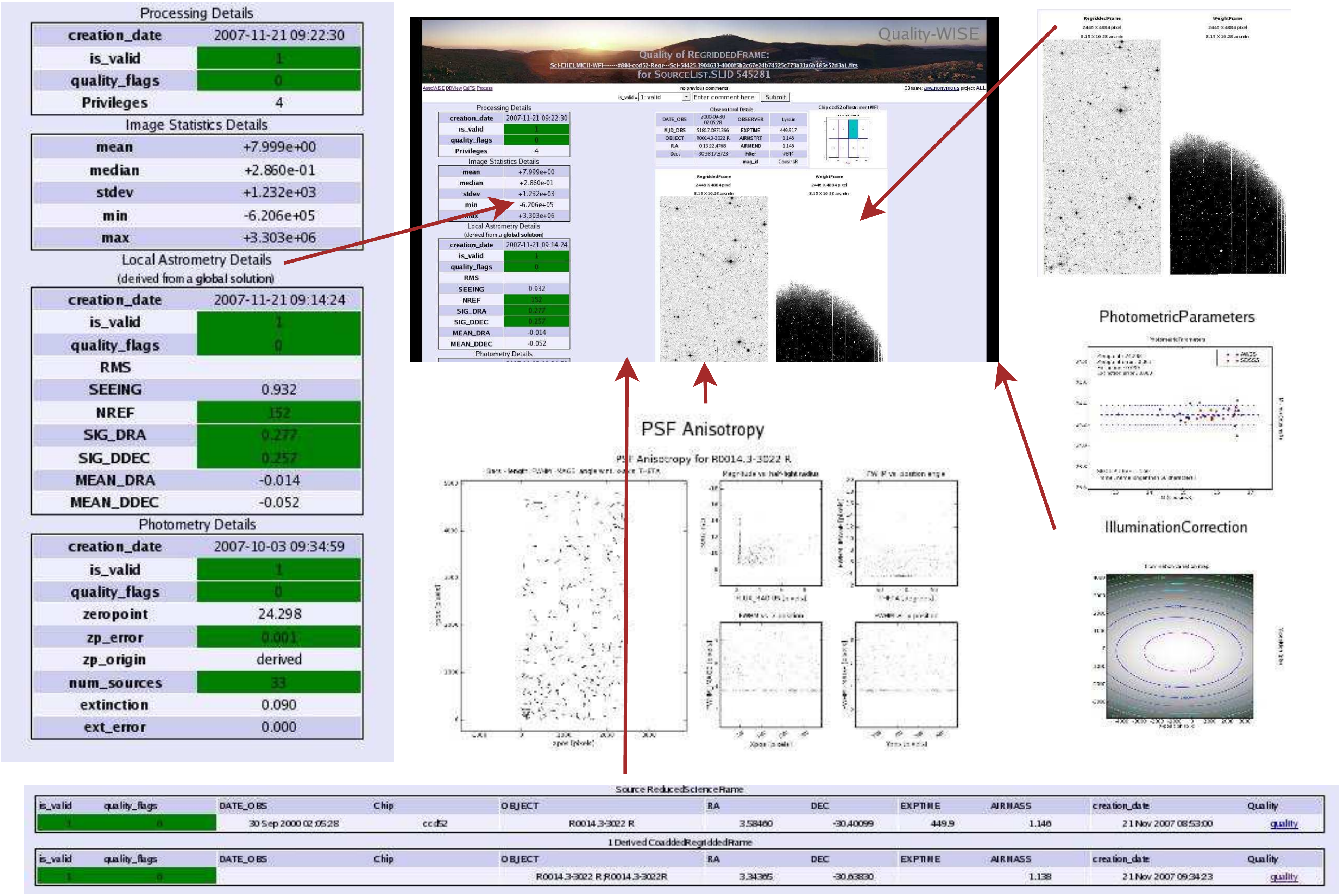}
\caption{The QualityWISE webservice bundles quality information for a single data object. An example page is shown partially, with arrows linking to zoom ins of the different sections of the webpage. 
See Quality Control paper in this issue \cite{qualitywise} for full discussion of this webservice and quality control in Astro-WISE in general.}
\label{f:qualityWise}
\end{figure}

The QualityWISE webservice provides an overview of the quality
assessment information (verdicts, inspection figures, numbers) for
human inspection, including links to the QualityWISE pages of data
items on which it depends.  Figure~\ref{f:qualityWise} gives an
impression of this service and for an in-depth discussion we refer to
the QualityWISE paper in this issue \cite{qualitywise}.  To zoom in on
quality issues, the user can use the webservice for database querying
and viewing (DBViewer\footnote{http://dbview.astro-wise.org}) which
provides links to the QualityWISE page of returned data items. At the
command-line the user can customize the quality assessment to
particular needs with maximum freedom. For example, batch scripts can
be used for mass quality control, with switches to an interactive mode
when needed.

\begin{figure}[ht]
\includegraphics[width=10.0cm]{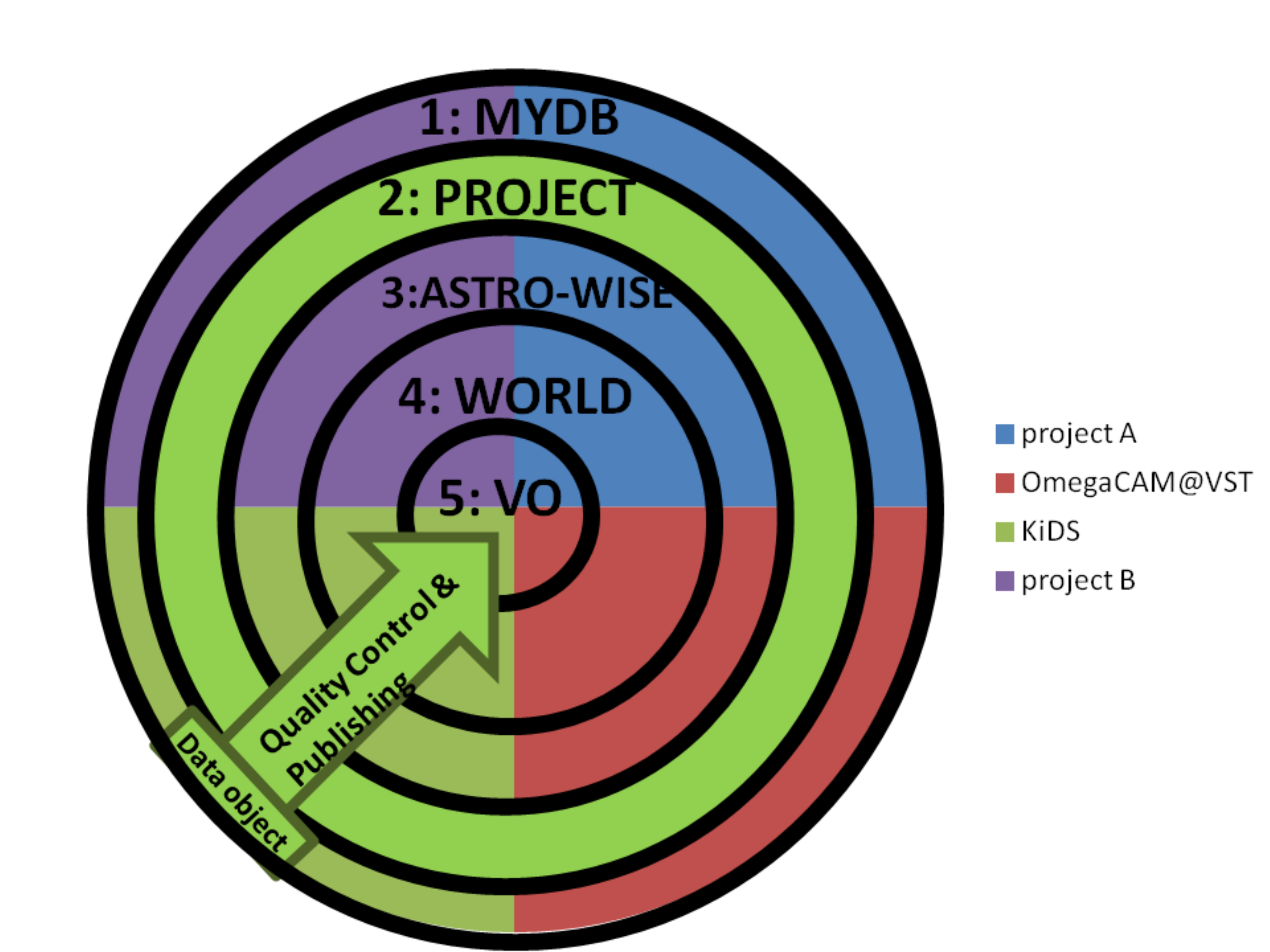}
\caption{Operational levels at which survey handling acts. This is
  configured by setting the minimum Privileges level in the user's
  Astro-WISE context. Each annulus represents a Privileges level. The
  operational level includes the annulus and all levels interior to
  it. Each color represents a project (KiDS in denoted in green). The
  baseline KiDS survey products reside at 2:PROJECT. These data can be
  accessed only by KiDS survey team members. Each team member can
  experiment to create improved versions of baseline products in
  her/his own level 1:MYDB, where data is only accessible by the
  single team member. If content, the member promotes the products to
  2:PROJECT to share them with the team. The KiDS project manager can
  publish baseline survey data from 2 to levels 3 to 5. Survey data at
  3:ASTRO-WISE can be accessed by all Astro-WISE users. At 4:WORLD,
  the data become accessible additionally to the astronomical
  community without an Astro-WISE account (anonymous users). At 5:VO,
  the data are accessible also from the Virtual Observatory.}
\label{f:contextGraphWithQC}
\end{figure}

For quality control it is important that baseline data products
readily distinghuished from experimental versions of data
products. The Privileges levels discussed earlier serve to keep such
versions apart. A KiDS team member can experiment to improve data
quality at the MYDB level. At this level, project data at all
Privileges levels is available, but the resulting products are only
visible to the team member.  Bad outcomes are discarded by
invalidating the data. Promising outcomes can be shared with the team
by publishing to the PROJECT level (see
Figure~\ref{f:contextGraphWithQC}). Fellow team members can then
inspect the data and provide feedback (e.g., using Comments objects in
Astro-WISE). The final verdict is set in the is\_valid attribute
(0=bad, 2=good). Upon team acceptance the survey object becomes
baseline and can be published higher up, eventually for delivery to
the outside world. Compromises in data quality identified only after
publishing can be handled adequately. For example, owing to the data
lineage provided by Astro-WISE, database queries with few lines of
code can isolate all data derived using a specific calibration
file. These can then easily be invalidated at all privileges levels.

\subsection{Survey calibration control}

Like science data, calibration data are represented as objects in
Astro-WISE. In addition to quality parameters, these objects carry a
creation date and editable timestamps that mark their validity
period. A request for a processing target generates a database query
that returns all good-quality calibration objects with a validity
period that covers the observation date. The newest good-quality
calibration object is then selected, following the survey handling
rule "newer is better". 

\begin{figure}[ht]
\includegraphics[width=11.9cm]{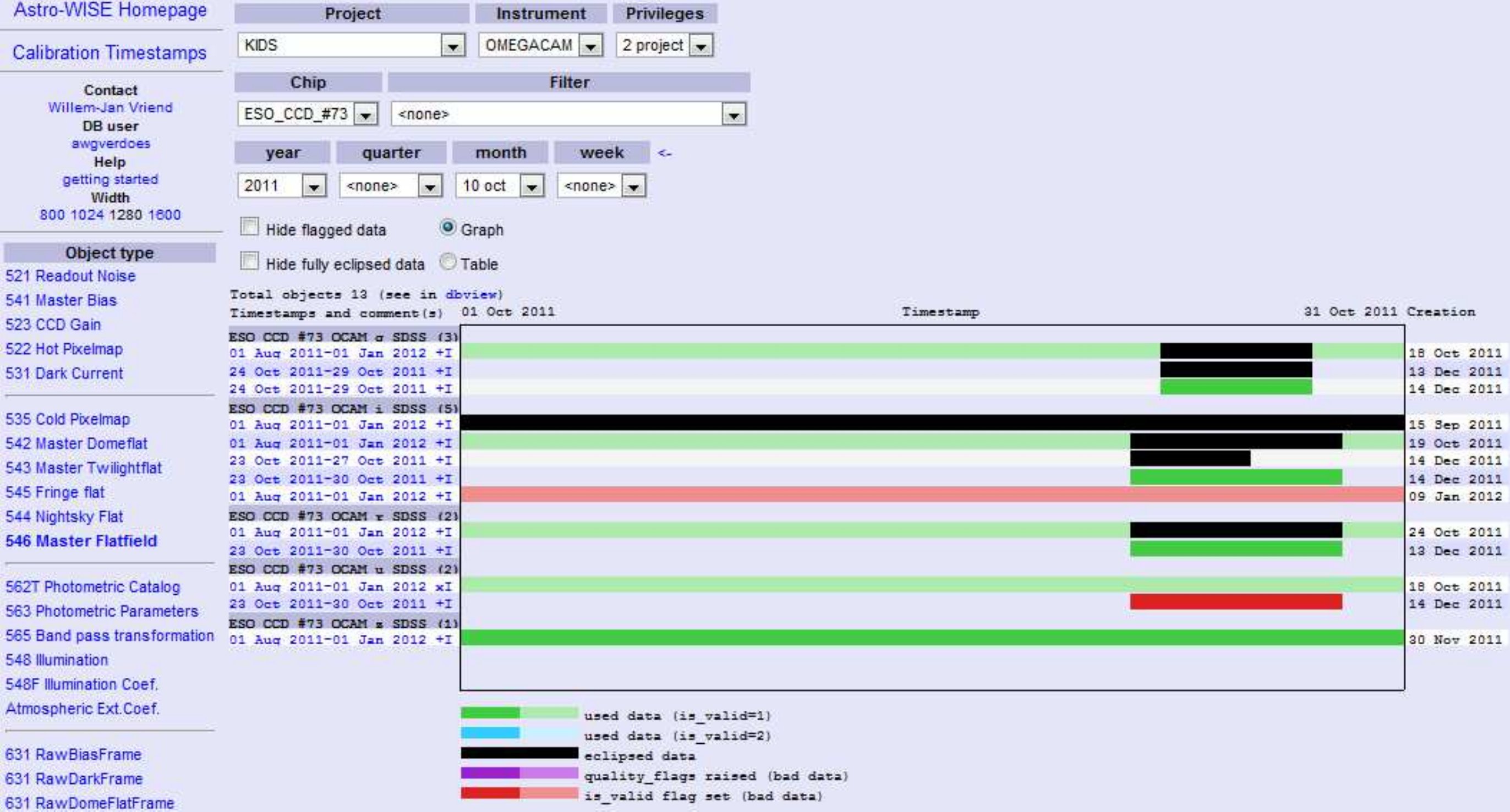}
\caption{Snapshot of the Calibration Time Stamp editor webservice
  (CalTS). All types of calibration data can be selected and their
  information graphically depicted. Horizontal bars show the time
  range validity of an object. The vertical stacking of the bars is
  ordered by creation date, with newer ones (green) eclipsing older
  versions (black). Other color codings depict the quality assessment
  verdicts. The validity time ranges and quality assessment parameters
  can be adjusted. See paper on User Interfaces in this issue
  \cite{userinterfaces} for technical details.}
\label{f:calts}
\end{figure}

The calibration scientists in the KiDS team, who are spread over
Europe, collaborate using the calibration control webservice
CalTS\footnote{http://calts.astro-wise.org/} to manipulate this
eclipsing of older calibrations by new ones (see Figure~\ref{f:calts}
and paper on user interfaces in this issue \cite{userinterfaces} for
details). As for science data, the Context parameters are used to
limit the survey calibration operations to the appropriate subset of
calibration data available in the system. As the survey progresses the
calibration scientists will build up a set of calibration objects with
a continuous time coverage. This build up of calibration data will
make subtle trends as a function of observational state (instrument
configuration, telescope position, atmospheric state) statistically
significant. Investigation of such trends can be done using the
Astro-WISE CLI in combination with scripts that can be as short as a
few lines of code.  The resulting deeper physical understanding of the
OmegaCAM instrument and the Paranal atmosphere will lead to better
calibrations which eclipse the older ones. The final result is that
the instrument plus atmosphere become continuously calibrated rather
than establishing calibrations on a per dataset basis.  This
continuous calibration coverage can be pooled with all Astro-WISE
users working on OmegaCAM data by publishing the data to privileges 3
or higher.

\section{From first reductions towards final data delivery}

KiDS survey operations have started officially on 15 October 2011.
The KiDS survey fields are currently being calibrated with initial
calibration data. Quality control on these calibration data and the
processed science data will lead to improvements of both the
calibration data objects as well as of the pipeline configuration and
methods. This in turn will allow the production of improved versions
of the initial survey data products.  Owing to the direct access to
data lineage in Astro-WISE it is straightforward to re-process only
the data objects that are affected by these changes. The resulting
high quality, calibrated survey data can then be used for the
production of advanced products, such as photometric redshifts, galaxy
morphometry, source variability analysis.

Following this approach the KiDS team will move from a 'quick-look'
version of the first survey products towards publishing a complete,
high quality, value-added KiDS Public Survey data set. All the time,
the team will benefit from Astro-WISE as its 'live archive', that
captures the accumulation of knowledge about OmegaCAM, VST and the
KiDS survey data over the years.

\begin{acknowledgements}
This work is financially supported by the Netherlands Research School for Astronomy (NOVA) and Target\footnote{www.rug.nl/target}. Target is supported by 
Samenwerkingsverband Noord Nederland, European fund for regional development, Dutch Ministry of economic affairs, Pieken in de Delta, Provinces of Groningen and 
Drenthe. Target operates under the auspices of Sensor Universe. The authors thank the referee for the constructive comments that helped to improve the paper.
\end{acknowledgements}

\end{document}